\documentstyle[11pt,newpasp,twoside]{article}
\begin{document}
\title{Self-Induced Formaton of Metal-Rich Globulars in Bulges?}
\author{Valery V. Kravtsov}
\affil{Sternberg Astronomical Institute, Universitetski pr. 13, Moscow,
119992, Russia}

\begin{abstract}
Taken together, key latest observations assume that (i) {\bf old} metal-rich
globular cluster populations (MRGCPs) in bulges were able to form due to
essentially increased, self(internally)-induced star formation rate (SFR) in
the hosts, while galaxy merging played an additional role; (ii) massive star
cluster populations (MSCPs) in irregulars may be young, less prominent
counterparts of the old MRGCPs in spheroids.
\end{abstract}

\section{Formation of MRGCPs in sheroids and MSCPs in irregulars}

Data on high redshift galaxies and QSOs, supermassive black holes, redshift
evolution of QSO emissivity, elemental abundances, etc. assume that the more
massive spheroid, the shorter timescale of its formation (e.g., Granato et al.
2001). The metallicity distribution functions (MDFs) for the disk stars of
the LMC and for the old red giants in the halos of elliptical NGC 5128 and
spiral M31 are virtually identical (Harris \& Harris 2001). Surprisingly
enough that the most probable metallicities of the MSCPs in the LMC and
other irregulars preferably fall (as metallicities of MRGCPs and of metal-rich
components of the MDFs do, too) between $0.004<z<0.008$, irrespective of
presence or absence of signs of interaction (Billett et al. 2002; de Grijs
et al. 2003, among others). In addition, for a sample of BCD galaxies Hopkins
et al. (2002) find a positive correlation between galaxy metallicity (oxygen
abundance) and SFR.
The above-mentioned imply that the formation of both the MRGCPs in spheroids
and MSCPs in irregulars may be preferably related to certain stage of the
host's (chemical?) evolution, at which the SFR in the hosts increases
essentially.

\acknowledgements

Many thanks are due to the organizers and the IAU for travel grant without
which my attendance at the GA would be impossible.


\begin{references}

\reference Billett, O. H., Hunter, D. A., \& Elmegreen, B. G. 2002, \aj,
123, 454
\reference de Grijs, R. et al. 2003, \mnras, 343, 1285
\reference Granato, G. L. et al. 2001, \mnras, 324, 757
\reference Harris, W. E., \& Harris, G. L. H. 2001, \aj, 122, 3065
\reference Hopkins, A. M., Schulte-Ladbeck, R. E., \& Drozdovsky, I. O.
2002, \aj, 124, 862
\end{references}
\end{document}